\documentclass[11pt,a4paper]{article}
\usepackage[margin=25mm]{geometry}
\usepackage[T1]{fontenc}
\usepackage[utf8]{inputenc}
\usepackage{lmodern}
\usepackage{microtype}
\usepackage{amsmath,amssymb,bm}
\usepackage{graphicx}
\usepackage{array}
\usepackage{xcolor}
\usepackage{cite}
\usepackage{placeins}
\usepackage[hidelinks]{hyperref}
\graphicspath{{./}}
\hypersetup{pdftitle={Non-Abelian Dirac oscillator in a uniform Yang-Mills background: exact spectrum and spin-isospin mixing},
  pdfauthor={Abdelmalek Boumali},pdfsubject={Revised manuscript, Paper No. 2606-014}}
\numberwithin{equation}{section}
\allowdisplaybreaks[2]

\newif\ifmarked
\ifdefined\RevisionMarked\markedtrue\else\markedfalse\fi
\newcommand{\rev}[1]{{\ifmarked\color{red}\else\color{black}\fi#1}}
\newcommand{\old}[1]{{\color{black}#1}}
\newenvironment{revised}{\color{black}}{}
\newenvironment{refereechange}{\ifmarked\color{red}\else\color{black}\fi}{}
\makeatletter
\AtBeginDocument{%
  \let\originaltagform@\tagform@
  \renewcommand{\tagform@}[1]{{\color{black}\originaltagform@{#1}}}%
}
\renewcommand{\@seccntformat}[1]{\textcolor{black}{\csname the#1\endcsname}\quad}
\renewcommand{\fnum@figure}{\textcolor{black}{\figurename\nobreakspace\thefigure}}
\makeatother

\newcommand{\KPR}{\mathcal{K}_{\mathrm{PR}}}
\newcommand{\Kzero}{K_0}
\newcommand{\LNA}{\Lambda_{\mathrm{NA}}}
\newcommand{\Hmin}{H_{\mathrm{min}}}
\newcommand{\HB}{H_B}
\newcommand{\Oeff}{\Omega_{\mathrm{eff}}}
\newcommand{\Drem}{\Delta_{\mathrm{rem}}}
\newcommand{\ii}{\mathrm{i}}
\newcommand{\ket}[1]{\lvert #1\rangle}
\newcommand{\doi}[1]{\href{https://doi.org/#1}{doi:\nolinkurl{#1}}}
\DeclareMathOperator{\tr}{tr}
\DeclareMathOperator{\spec}{spec}
\DeclareMathOperator{\atanTwo}{atan2}

\title{Non-Abelian Dirac oscillator in a uniform Yang--Mills background:\\
\rev{exact spectrum and} spin--isospin mixing}
\author{Abdelmalek Boumali$^{1,*}$\\[0.3em]
{\small ORCID: 0000-0003-2552-0427}\\[0.2em]
{\small arXiv: 2605.30403}\\[0.3em]
{\small $^1$Laboratory of Theoretical and Applied Physics,}\\
{\small Echahid Cheikh Larbi Tebessi University, Tebessa, Algeria}\\
{\small ROR: 00tnac320}\\
{\small $^*$Corresponding author: \texttt{boumali.abdelmalek@gmail.com}}}
\date{}

\begin{document}
\maketitle

\begin{abstract}
\rev{We study} \old{a planar Dirac oscillator} \rev{in a uniform gauge background with} \old{an Abelian magnetic} \rev{component and two} \old{non-Abelian} \rev{amplitudes. The noncommuting spatial potentials generate a chromomagnetic curvature along $T^3=\tau^3/2$, while the temporal potential generates chromoelectric components proportional to their product. We derive the second-order stationary identity without discarding the isospin--orbit, diamagnetic, or scalar-potential terms. The} \old{first-order} \rev{Hamiltonian, with minimal coupling, a Pauli interaction, or both, admits a conserved oscillator--spin--isospin operator and reduces to finite blocks. We obtain the exact secular equations, the exceptional lowest signed levels, and the leading avoided-crossing gaps. Minimal coupling produces transverse mixing already when the scalar amplitude vanishes; a gap proportional to the product of the amplitudes instead occurs at resonance in the pure Pauli model. We also solve a separately defined field-strength comparison operator. Its spin-dependent backbone does not commute with the transverse internal coupling, so its spectrum is not obtained by adding the isolated internal eigenvalues to the backbone energies. The mixed branches have quadratic weak-coupling shifts away from degeneracy. We distinguish gauge invariants from internal-basis labels and give conditional, spectrally resolved parameter examples, without claiming demonstrated graphene or atomic realizations.}
\end{abstract}


\section{Introduction}\label{sec:introduction}
The Dirac oscillator is a paradigmatic exactly solvable relativistic model. It is obtained by introducing a linear non-minimal coupling in the Dirac equation\rev{, and it} combines a discrete oscillator-like spectrum with a transparent spin structure \rev{\cite{Moshinsky1989,Benitez1990}}. Its two-dimensional version is useful as an analytical benchmark for magnetic confinement, \rev{graphene-related applications,} and relativistic quantum simulations \rev{\cite{Bermudez2008,Gerritsma2010,Franco2013,Belouad2016,Boumali2015}. The oscillator substitution is}
\begin{refereechange}
\begin{equation}
\bm p\longrightarrow\bm p-\ii m\omega\beta_D\bm r,\qquad \beta_D=\gamma^0.
\label{eq:oscillator}
\end{equation}
A spatially uniform non-Abelian connection can have nonzero curvature because its components need not commute. The uniform background of Dossa and Avossevou \cite{Dossa2020} provides a concrete setting in which to investigate this mechanism.

The aligned construction of Ref.~\cite{BoumaliGarah} motivates the \old{internal-Zeeman} coupling considered here. Its internal generator is explicitly $T^3=\tau^3/2$, selected by the commutator of the two spatial connection components. An additional temporal component produces electric-type curvatures in the transverse internal directions. However, identifying these curvatures does not by itself determine the Dirac spectrum: the kinetic connection and scalar-potential terms must also be retained\old{.}

Our primary object is therefore the first-order Hamiltonian. Its exact finite-block solution separates the minimal and Pauli-coupling mechanisms and retains the signed lowest states. A second object, $\KPR$, is kept only as an independently defined phenomenological comparison model that isolates a Hermitian spin--field-strength interaction. It is neither the square of the specified minimally coupled Hamiltonian nor a controlled truncation of its exact stationary reduction. Even within this comparison model, \old{the} transverse internal term cannot be diagonalized independently of the backbone spin term. This distinction changes both the energy formula and the interpretation of the internal singlet/triplet labels\old{.}

\old{Section~\ref{sec:model}} fixes the conventions\old{. Section}~\ref{sec:firstorder} gives the exact first-order spectrum and resonant mixing\old{. Section}~\ref{sec:comparison} derives the complete comparison spectrum. Section~\ref{sec:gauge} discusses gauge and basis dependence, and Sec.~\ref{sec:regimes} gives conditional parameter regimes. Appendix~\ref{app:reduction} supplies the operator-level second-order reduction, and Appendix~\ref{app:secular} details the spectral derivation\old{.}
\end{refereechange}

\section{Model, conventions, and field strength}\label{sec:model}
We use natural units, \rev{$\hbar=c=1$. Spacetime indices are $\mu,\nu=0,1,2$, spatial indices are $i,j=1,2$, and internal indices are $a,b,c=1,2,3$. The internal tensor product is}
\begin{revised}
\begin{equation}
\Psi\in\mathbb{C}^{2}_{\mathrm{spinor}}\otimes\mathbb{C}^{2}_{\mathrm{iso}},
\qquad T^a=\frac{\tau^a}{2},\qquad
\old{[}T^a,T^b\old{]}=\ii\epsilon^{abc}T^c.
\label{eq:internal}
\end{equation}
\old{The} orbital factor $L^2(\mathbb{R}^2)$ is implicit. In a material realization, ``spinor'' can be a sublattice pseudospin rather than physical spin. We suppress tensor-product identities when no ambiguity arises. We take $g>0$ as a coupling convention, $m\geq0$, and real $\beta,\rho,\omega$. Formulas containing $1/m$ require $m>0$\old{.}

We fix the planar representation and spacetime metric as
\begin{equation}
\gamma^0=\sigma^3,\quad \gamma^1=\ii\sigma^2,\quad \gamma^2=\ii\sigma^1,
\qquad \alpha^1=\sigma^1,\quad \alpha^2=-\sigma^2,\quad
\eta=\operatorname{diag}(1,-1,-1).
\label{eq:gamma}
\end{equation}
The connection and covariant derivative are
\begin{equation}
\mathcal{A}_\mu=eA_\mu^{(0)}\mathbb{I}_2+gA_\mu^aT^a,
\qquad \mathcal{D}_\mu=\partial_\mu+\ii\mathcal{A}_\mu.
\label{eq:connection}
\end{equation}
This uses the $\mathfrak{u}(1)\oplus\mathfrak{su}(2)$ decomposition of the matrix Lie algebra and standard Yang--Mills conventions \cite{YangMills1954}. With this convention,
\begin{align}
F_{\mu\nu}^{(0)}&=\partial_\mu A_\nu^{(0)}-\partial_\nu A_\mu^{(0)},\label{eq:abelian-curvature}\\
F_{\mu\nu}^a&=\partial_\mu A_\nu^a-\partial_\nu A_\mu^a-g\epsilon^{abc}A_\mu^bA_\nu^c.\label{eq:nonabelian-curvature}
\end{align}
Changing the sign convention for the covariant derivative requires changing the potential/coupling conventions consistently; it is not a change of the field strength alone.

\old{The Abelian} field \old{is chosen in the symmetric gauge,}
\begin{equation}
A_0^{(0)}=0,\qquad A_i^{(0)}=-\frac{B}{2}\epsilon_{ij}x_j,
\qquad F_{12}^{(0)}=B,\qquad \epsilon_{12}=1.
\label{eq:symmetric-gauge}
\end{equation}
\old{The non-Abelian} background is \cite{Dossa2020}
\begin{equation}
A_1^2=-\beta,\qquad A_2^1=\beta,\qquad A_0^3=\rho,
\label{eq:ansatz}
\end{equation}
with all other connection components zero\old{. Direct substitution gives}
\begin{equation}
F_{12}^3=-g\beta^2,\qquad F_{01}^1=F_{02}^2=-g\beta\rho.
\label{eq:field-strength}
\end{equation}
The remaining components follow by antisymmetry or vanish. This is an externally prescribed Yang--Mills background; a source-free field equation is not imposed. Its connection is spatially constant in \old{the non-Abelian} sector, not in the Abelian symmetric gauge.

\begin{refereechange}
The generator selected by the spatial commutator is
\begin{equation}
gF_{12}^aT^a=-g^2\beta^2T^3,\qquad T^3=\tau^3/2.
\label{eq:magnetic-generator}
\end{equation}
Thus $T^3$ produces the magnetic \old{internal-Zeeman} structure. At $\rho=0$ the curvature is purely chromomagnetic, but the spatial potentials still contain the noncommuting generators $T^1,T^2$. Consequently the minimally coupled Hamiltonian need not be diagonal in $T^3$ even in this aligned-curvature limit.
\end{refereechange}

Define
\begin{equation}
\Oeff=m\omega+\frac{eB}{2},\quad
\Pi_1=p_x-\Oeff y,\quad \Pi_2=p_y+\Oeff x,\quad
[\Pi_1,\Pi_2]=-2\ii\Oeff.
\label{eq:pi}
\end{equation}
All discrete ladder constructions below assume $\Oeff>0$; we denote it by $\Omega$. For a general sign the squared Abelian backbone is $m^2+2|\Oeff|(2n+1)-2\Oeff s$, $s=\pm1$. At negative $\Oeff$ the ladder orientation and unpaired spinor state reverse. At zero $\Oeff$ the discrete oscillator construction is unavailable. These cases are not included in the finite-block indexing used below.

\section{Exact first-order Hamiltonian and spectrum}\label{sec:firstorder}
\subsection{Minimal and Pauli couplings}\label{sec:couplings}
\begin{refereechange}
The minimally coupled stationary Dirac equation is $\Hmin\psi=E\psi$, with
\begin{align}
\Hmin&=\alpha^j(p_j+\mathcal{A}_j-\ii m\omega\beta_Dx_j)+\beta_Dm+\mathcal{A}_0\notag\\
&=\HB+g\rho T^3-g\beta\Gamma,\label{eq:hmin}\\
\HB&=\sigma^1\Pi_1-\sigma^2\Pi_2+m\sigma^3,
\qquad\Gamma=\sigma^1T^2+\sigma^2T^1.\label{eq:hb-gamma}
\end{align}
The off-diagonal term in this Hamiltonian is already linear in $\beta$ at $\rho=0$. It must not be confused with the product-dependent curvature term of Eq.~\eqref{eq:field-strength}.

To compare with an anomalous Pauli-type interaction, define explicitly
\begin{equation}
V_P=\frac{\kappa}{4m}\beta_D\sigma^{\mu\nu}gF_{\mu\nu}^aT^a
=\frac{\kappa g^2}{2m}\bigl(\beta^2T^3+\beta\rho\Gamma\bigr),
\qquad\sigma^{\mu\nu}=\frac{\ii}{2}[\gamma^\mu,\gamma^\nu].
\label{eq:pauli}
\end{equation}
Repeated antisymmetric indices in this equation include both orders. This is a specified Hamiltonian convention with real $\kappa$ and $m>0$. It follows from $\beta_D\sigma^{12}=-\mathbb{I}_2$, $\beta_D\sigma^{01}=-\sigma^2$, and $\beta_D\sigma^{02}=-\sigma^1$\old{.}

All three coupling choices have the common form
\begin{equation}
H=\HB+2rT^3-k\Gamma.
\label{eq:common-h}
\end{equation}
Their parameters must be distinguished:
\begin{equation}
\begin{array}{c|cc}
\text{coupling}&r&k\\\hline
\text{minimal only}&g\rho/2&g\beta\\
\text{Pauli only}&\kappa g^2\beta^2/(4m)&-\kappa g^2\beta\rho/(2m)\\
\text{minimal plus Pauli}&g\rho/2+\kappa g^2\beta^2/(4m)&g\beta-\kappa g^2\beta\rho/(2m)
\end{array}
\label{eq:coupling-table}
\end{equation}
At $\rho=0$, the pure-Pauli row has the same aligned $T^3$ splitting structure as Ref.~\cite{BoumaliGarah}. In that paper the transverse connection is written as $A_1^1=A_2^2=\eta$ and the Pauli contribution is a rigid $T^3$ shift. The two transverse-potential conventions are related by a constant internal rotation about the third axis. That paper absorbs $\kappa$ into its symbol $\zeta$, whereas here $\zeta=g^2\beta^2/(2m)$ and $\kappa$ is kept explicit. The pure Pauli comparison omits both non-Abelian minimal terms while retaining the Abelian oscillator $\HB$. It is a separately specified fixed-frame effective Hamiltonian, not the minimally gauged theory with terms silently removed. Its covariance qualifications are given in Sec.~\ref{sec:gauge}. For every real $r,k$\old{, Eq.}~\eqref{eq:common-h} is a bounded Hermitian internal perturbation of the self-adjoint oscillator Hamiltonian, so its energy spectrum is real without a parameter-dependent square-root positivity condition.
\end{refereechange}

\subsection{Conserved quantity and finite blocks}\label{sec:blocks}
\begin{refereechange}
Introduce $\Pi_\pm=\Pi_1\pm \ii\Pi_2$ and
\begin{equation}
\Pi_-=2\sqrt{\Omega}\,b,\quad \Pi_+=2\sqrt{\Omega}\,b^\dagger,
\qquad[b,b^\dagger]=1,\qquad T^\pm=T^1\pm \ii T^2.
\label{eq:ladder}
\end{equation}
Then
\begin{equation}
H=\begin{pmatrix}
m+2rT^3&2\sqrt{\Omega}\,b^\dagger+\ii kT^+\\
2\sqrt{\Omega}\,b-\ii kT^-&-m+2rT^3
\end{pmatrix},\qquad
Q=b^\dagger b+T^3-\frac{\sigma^3}{2},\quad[H,Q]=0.
\label{eq:q-conserved}
\end{equation}
\old{The} second, guiding-centre oscillator commutes with $H$ and supplies the usual Landau degeneracy on the full plane. We keep one such spectator sector fixed. States in a $Q=q\geq1$ sector are ordered as
\begin{equation}
\ket{q,+,\uparrow},\quad\ket{q+1,-,\uparrow},\quad
\ket{q-1,+,\downarrow},\quad\ket{q,-,\downarrow},
\label{eq:basis}
\end{equation}
where the entries are oscillator number, isospin sign, and spinor component. Multiplying the second and fourth basis states by $-\ii$ gives the real block
\begin{equation}
H_q=\begin{pmatrix}
m+r&0&a_q&k\\
0&m-r&0&a_{q+1}\\
a_q&0&-m+r&0\\
k&a_{q+1}&0&-m-r
\end{pmatrix},\qquad a_q=2\sqrt{\Omega q}.
\label{eq:hq}
\end{equation}
At $q=0$ the third state is absent and
\begin{equation}
H_0=\begin{pmatrix}
m+r&0&k\\0&m-r&a_1\\k&a_1&-m-r
\end{pmatrix},\qquad H_{-1}=(m-r).
\label{eq:exceptional-blocks}
\end{equation}
There are four states for every $q\geq1$, three for $q=0$, and one for $q=-1$, per spectator sector. This signed state counting is essential when interpreting a squared equation.
\end{refereechange}

\subsection{Secular equations and closed limits}\label{sec:secular}
\begin{refereechange}
Set $\varepsilon_q=\sqrt{m^2+4\Omega q}$. The exact characteristic equation for $q\geq1$ is
\begin{equation}
\bigl[(E-r)^2-\varepsilon_q^2\bigr]\bigl[(E+r)^2-\varepsilon_{q+1}^2\bigr]
=k^2(E-r+m)(E+r-m).
\label{eq:quartic}
\end{equation}
The exceptional equations are
\begin{align}
(E-r-m)\bigl[(E+r)^2-\varepsilon_1^2\bigr]&=k^2(E+r-m),\qquad q=0,\label{eq:cubic}\\
E&=m-r,\qquad q=-1.\label{eq:isolated}
\end{align}
These equations follow directly from the first-order blocks; no energy-sign choice or root exclusion is left implicit.

For $k=0$ and $q\geq1$, the roots are $r\pm\varepsilon_q$ and $-r\pm\varepsilon_{q+1}$. For the Abelian limit $r=k=0$, their union with the exceptional sectors is the familiar pair of isospin copies of \old{the} planar Dirac-oscillator ladder: an unpaired $E=m$ in each copy and $\pm\sqrt{m^2+4\Omega N}$ for integers $N\geq1$. Setting $B=0$ gives the pure oscillator; setting $\omega=0$ gives the relativistic Landau problem in the chosen orientation.

At $r=0$, define
\begin{equation}
X_{q,\pm}=m^2+2\Omega(2q+1)+\frac{k^2}{2}
\pm\sqrt{4\Omega^2+2\Omega(2q+1)k^2+\frac{k^4}{4}}.
\label{eq:xq}
\end{equation}
For $q\geq1$ all four signed roots $E=\pm\sqrt{X_{q,+}}$, $\pm\sqrt{X_{q,-}}$ occur. At the lowest sectors the correct result is instead
\begin{equation}
\begin{aligned}
q=0:&\quad E=m,\quad E=\pm\sqrt{m^2+4\Omega+k^2},\\
q=-1:&\quad E=m.
\end{aligned}
\label{eq:lowest-rzero}
\end{equation}
For $m>0$ there is no extra $E=-m$ state in these sectors. At $m=0$ the two listed zero levels have the multiplicities already specified. The condition $r=0$ equals $\rho=0$ for minimal coupling, but not for the general Pauli-coupled family. For minimal coupling the isolated state at arbitrary $\rho$ is $m-g\rho/2$, not always $m$.

Expanding \old{Eq.}~\eqref{eq:xq} at fixed $q$ gives
\begin{equation}
X_{q,-}=m^2+4\Omega q-qk^2+O(k^4/\Omega),\quad
X_{q,+}=m^2+4\Omega(q+1)+(q+1)k^2+O(k^4/\Omega).
\label{eq:xq-expansion}
\end{equation}
\old{The} expansion is nonuniform in $q$; a sufficient smallness condition is $(2q+1)k^2/\Omega\ll1$. It must not be extrapolated to infer a negative squared energy of the exact Hermitian first-order model. Figure~\ref{fig:aligned} displays its actual positive roots.
\end{refereechange}

\begin{figure}[tbp]
\centering
\includegraphics[width=0.91\textwidth]{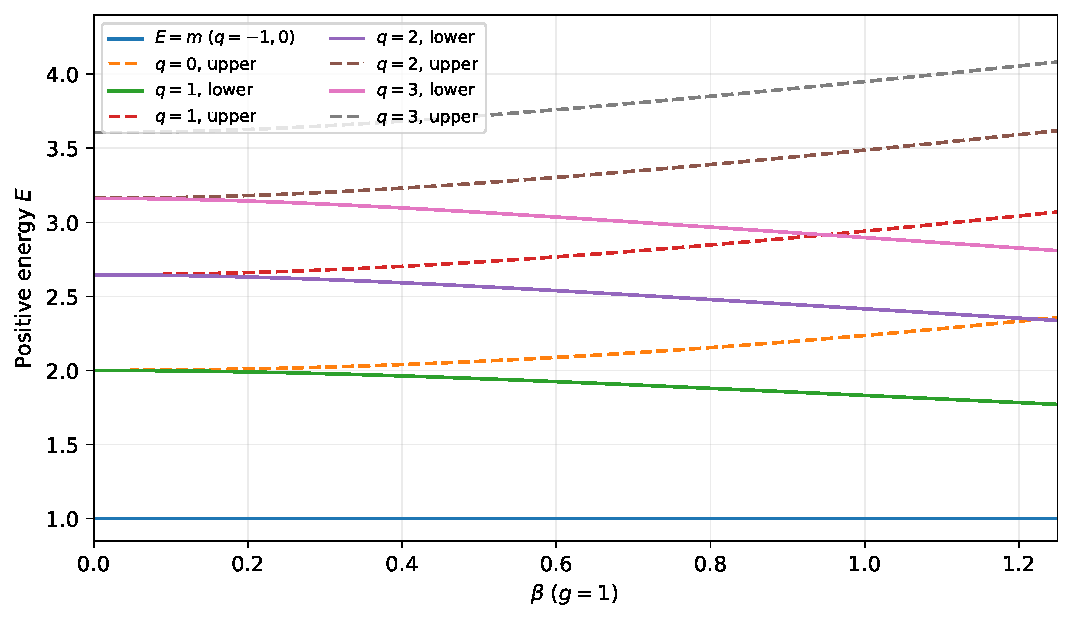}
\caption{\rev{Exact positive first-order levels for minimal coupling at $\rho=0$, $m=g=1$, $\Omega=0.75$. The unpaired levels from $q=-1$ and $q=0$ coincide at $E=m$. Other levels have $q$-dependent shifts; the internal field-strength eigenvalues are not added to the backbone spectrum. Each curve is obtained by diagonalizing the corresponding complete first-order block.}}
\label{fig:aligned}
\end{figure}

\subsection{Nonresonant shifts and avoided crossings}\label{sec:crossings}
Away from a degeneracy within a fixed $Q$ sector, the positive level $r+\varepsilon_q$ has shift
\begin{equation}
\delta E=\frac{k^2(\varepsilon_q+m)(\varepsilon_q+2r-m)}
{2\varepsilon_q[(\varepsilon_q+2r)^2-\varepsilon_{q+1}^2]}+O(k^4),
\label{eq:shift}
\end{equation}
where the remainder is understood at fixed nonzero energy denominators. The relevant perturbative control is coupling relative to those denominators, not merely $|k|$ relative to the energy itself.

The two positive uncoupled levels meet at the bare resonance
\begin{equation}
2r=\varepsilon_{q+1}-\varepsilon_q.
\label{eq:resonance}
\end{equation}
Their projected matrix element is
\begin{equation}
M_q=k\sqrt{\frac{\varepsilon_q+m}{2\varepsilon_q}}\sqrt{\frac{\varepsilon_{q+1}-m}{2\varepsilon_{q+1}}}.
\label{eq:mq}
\end{equation}
When $|k|$ is small relative to the remote-state separation $\Drem$, the gap at this resonance is
\begin{equation}
\Delta_q=2|M_q|+O(k^2/\Drem)
=|k|\sqrt{\frac{(\varepsilon_q+m)(\varepsilon_{q+1}-m)}{\varepsilon_q\varepsilon_{q+1}}}+O(k^2/\Drem).
\label{eq:gap}
\end{equation}
Equal-weight combinations of the two uncoupled levels are eigenstates only in the leading resonant two-level approximation. Remote-state admixtures and \old{second-order} diagonal shifts modify the exact eigenvectors and the precise position of the minimum gap at finite $k$. Crossings between different $Q$ sectors are not avoided by this Hamiltonian.

For minimal coupling, $k=g\beta$ and $2r=g\rho$: the scalar potential tunes the crossing, while the spatial connection opens a gap already present as a matrix element at $\rho=0$. For the pure Pauli model, $2r=\kappa g^2\beta^2/(2m)$ and $k=-\kappa g^2\beta\rho/(2m)$. In the latter case a resonant gap is linear in $|\beta\rho|$ to leading order. This does not imply a uniform product-linear shift of every nondegenerate level. Figures~\ref{fig:minimal-crossing} and~\ref{fig:pauli-crossing} distinguish the two mechanisms.

\begin{figure}[tbp]
\centering
\includegraphics[width=0.91\textwidth]{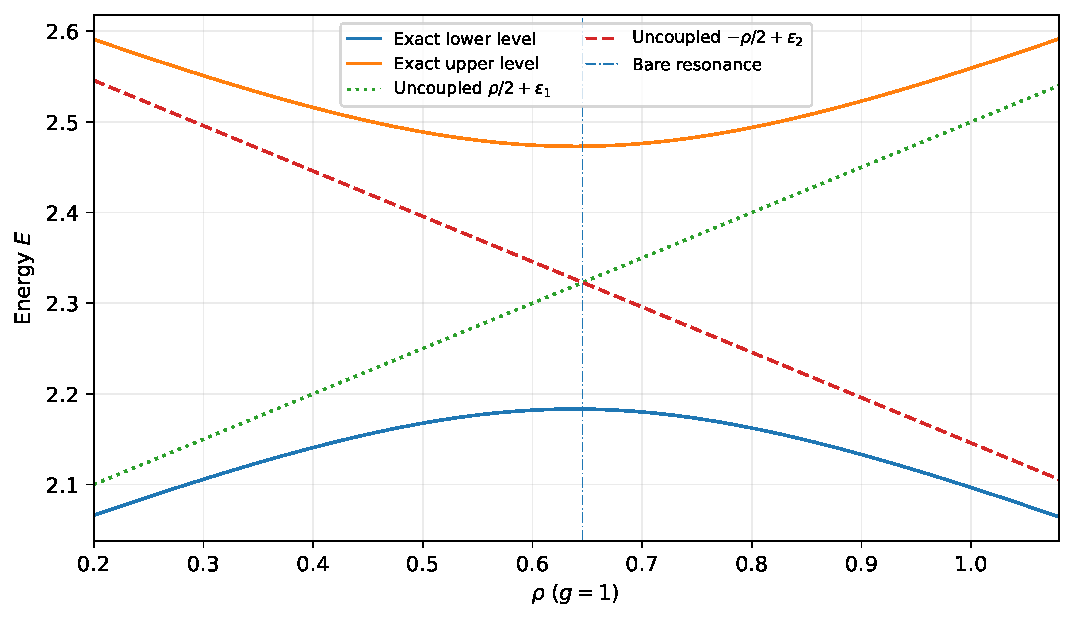}
\caption{\rev{Minimal-coupling avoided crossing in the $q=1$ sector, with $m=g=1$, $\Omega=0.75$, and $\beta=0.3$. The two exact positive levels are compared with the uncoupled references $\rho/2+\varepsilon_1$ and $-\rho/2+\varepsilon_2$. The vertical line is the bare resonance $\rho=\sqrt{7}-2$; it is not asserted to be the exact finite-coupling minimum-gap position.}}
\label{fig:minimal-crossing}
\end{figure}
\begin{figure}[tbp]
\centering
\includegraphics[width=0.91\textwidth]{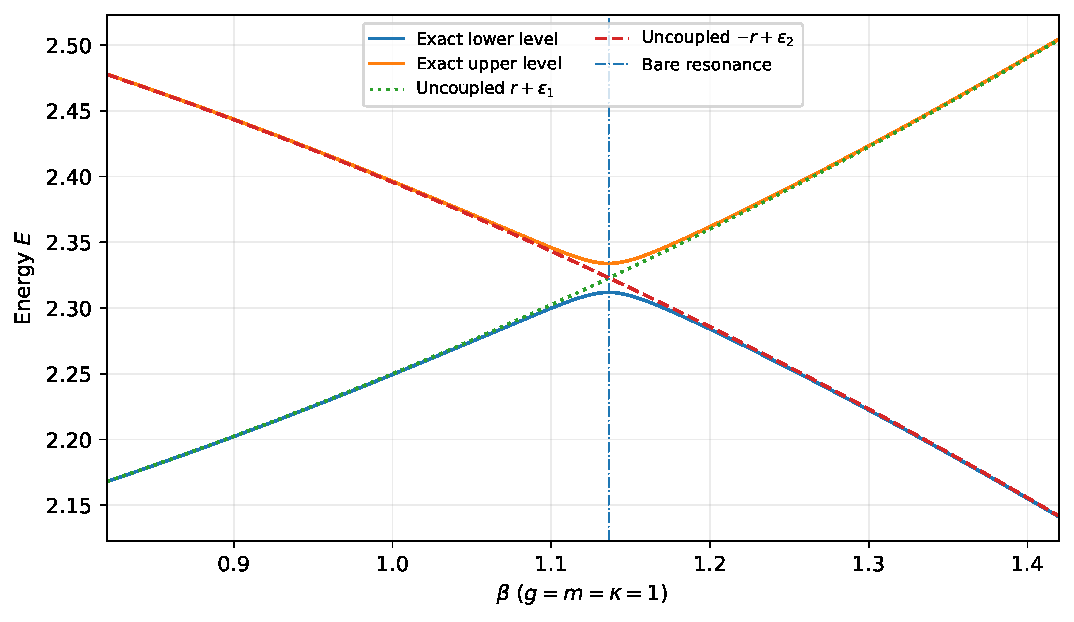}
\caption{\rev{Pure Pauli model in the $q=1$ sector, with $m=g=\kappa=1$, $\Omega=0.75$, and $\rho=0.04$. The non-Abelian minimal terms are absent. Here $r=\beta^2/4$ and $k=-\beta\rho/2$. The bare resonance occurs at $\beta=\sqrt{2(\sqrt{7}-2)}$. The weak transverse term opens a product-dependent gap near this crossing, rather than imposing a product-linear shift away from resonance.}}
\label{fig:pauli-crossing}
\end{figure}
\FloatBarrier

\section{Internal field-strength operator and comparison model}\label{sec:comparison}
\subsection{An internal diagnostic, not an additive Dirac spectrum}\label{sec:diagnostic}
For $m>0$, define the Hermitian field-strength diagnostic
\begin{equation}
\LNA=\frac{g}{2m}(-\sigma^3F_{12}^a+\sigma^iF_{0i}^a)T^a
=\frac{g^2}{2m}\Bigl[\beta^2\sigma^3T^3-\beta\rho(\sigma^1T^1+\sigma^2T^2)\Bigr].
\label{eq:lambda-na}
\end{equation}
Its magnetic part agrees with the corresponding term of the exact second-order reduction. Its electric part is a Hermitian diagnostic chosen for comparison, not the electric commutator of that reduction. In particular, $\sigma^1T^1+\sigma^2T^2$ is not the \old{first-order} operator $\Gamma$.

In the ordered spinor--isospin basis $\{\ket{+,+},\ket{+,-},\ket{-,+},\ket{-,-}\}$,
\begin{equation}
\LNA=\frac{g^2}{4m}\begin{pmatrix}
\beta^2&0&0&0\\0&-\beta^2&-2\beta\rho&0\\
0&-2\beta\rho&-\beta^2&0\\0&0&0&\beta^2
\end{pmatrix}.
\label{eq:lambda-matrix}
\end{equation}
The aligned states $\ket{+,+},\ket{-,-}$ have the same internal eigenvalue, while the antisymmetric and symmetric combinations are
\begin{equation}
\ket{S}=\frac{\ket{+,-}-\ket{-,+}}{\sqrt{2}},\qquad
\ket{T_0}=\frac{\ket{+,-}+\ket{-,+}}{\sqrt{2}}.
\label{eq:singlet-triplet}
\end{equation}
Their eigenvalues are
\begin{align}
\lambda_\parallel&=\frac{g^2\beta^2}{4m}\quad\text{(multiplicity two)},\notag\\
\lambda_S&=-\frac{g^2\beta(\beta-2\rho)}{4m},\qquad
\lambda_{T_0}=-\frac{g^2\beta(\beta+2\rho)}{4m},\label{eq:internal-eigenvalues}\\
\lambda_S-\lambda_{T_0}&=\frac{g^2\beta\rho}{m}.
\label{eq:internal-splitting}
\end{align}
These are eigenvalues of $\LNA$ only. Equation~\eqref{eq:internal-splitting} is an internal-operator splitting and is not the generic energy splitting of either $H$ or $\KPR$. In a common-axis basis, $J^a=\sigma^a/2+T^a$ satisfies $[\LNA,J^2]=0$: anisotropy lifts the triplet degeneracy, rather than destroying the singlet/triplet algebraic decomposition. These names nevertheless depend on how the two internal bases are identified (Sec.~\ref{sec:gauge}). Figure~\ref{fig:internal} shows this internal algebra without treating it as a full energy spectrum.

\begin{figure}[tbp]
\centering
\includegraphics[width=0.91\textwidth]{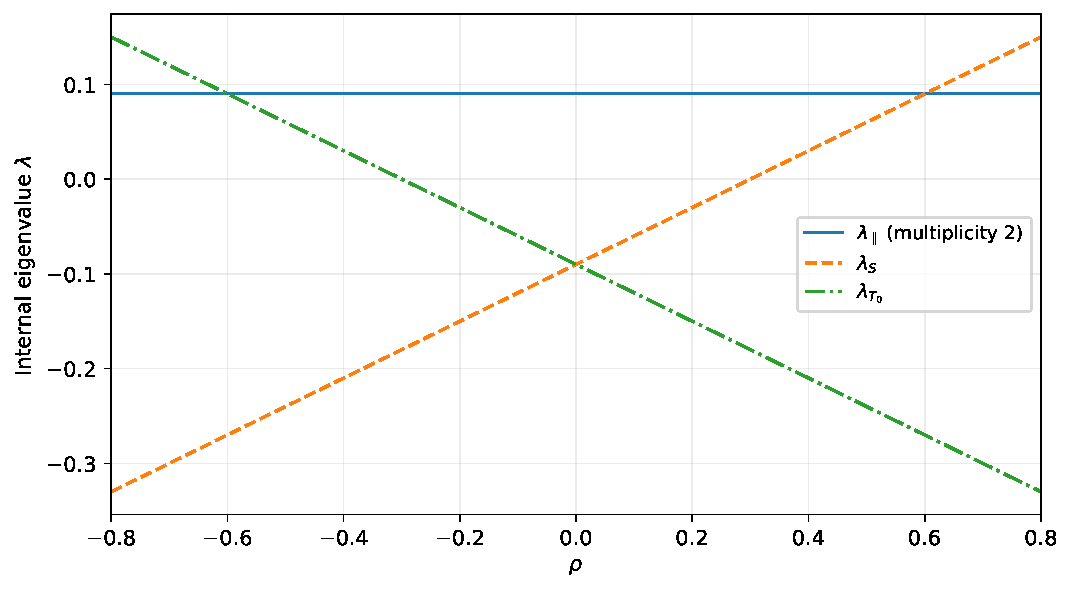}
\caption{\rev{Eigenvalues of the isolated internal diagnostic $\LNA$, with $m=g=1$ and $\beta=0.6$. The aligned eigenvalue has multiplicity two. The linear separation of $\lambda_S$ and $\lambda_{T_0}$ is an internal-operator result, not a generic energy separation of the first-order Hamiltonian or the full comparison operator.}}
\label{fig:internal}
\end{figure}

\begin{refereechange}
\subsection{Full diagonalization of the comparison operator}\label{sec:full-comparison}
The phenomenological second-order eigenvalue problem retained for comparison is
\begin{equation}
\KPR\psi=E^2\psi,\qquad \KPR=\Kzero+2m\LNA,\qquad
\Kzero=\Pi_1^2+\Pi_2^2+m^2-2\Omega\sigma^3.
\label{eq:kpr}
\end{equation}
Despite the constancy of $\LNA$, it does not commute with $\Kzero$ in general:
\begin{equation}
\old{[}\Kzero,\LNA\old{]}=\frac{2\ii\Omega g^2\beta\rho}{m}(\sigma^2T^1-\sigma^1T^2).
\label{eq:noncommutation}
\end{equation}
The orbital part commutes; the backbone spin term does not. Therefore adding $2m\lambda_k$ to a backbone squared energy with an independent $s=\pm1$ label is invalid when $\Omega\beta\rho\ne0$. The $S,T_0$ states are not $\sigma^3$ eigenstates\old{.}

Only the orbital oscillator separates. Define
\begin{equation}
C_n=m^2+2\Omega(2n+1),\quad A=\frac{g^2\beta^2}{2},\quad
C=g^2\beta\rho,\quad R=\sqrt{4\Omega^2+C^2}.
\label{eq:comparison-parameters}
\end{equation}
At fixed $n\geq0$ the complete internal matrix is
\begin{equation}
\left.\KPR\right|_n=\begin{pmatrix}
C_n-2\Omega+A&0&0&0\\
0&C_n-2\Omega-A&-C&0\\
0&-C&C_n+2\Omega-A&0\\
0&0&0&C_n+2\Omega+A
\end{pmatrix}.
\label{eq:comparison-matrix}
\end{equation}
Its four squared-energy branches, with no extra independent spin label, are
\begin{equation}
\begin{aligned}
E_{n,++}^2&=C_n-2\Omega+A,\\
E_{n,--}^2&=C_n+2\Omega+A,\\
E_{n,\mathrm{mix},\pm}^2&=C_n-A\pm\sqrt{4\Omega^2+g^4\beta^2\rho^2}.
\end{aligned}
\label{eq:comparison-spectrum}
\end{equation}
The notation $E^2$ specifies this second-order spectral problem. Writing both formal roots $\pm\sqrt{E^2}$ is a convention for this model, not evidence that either sign satisfies the minimally coupled first-order equation.

For the lower mixed eigenvalue let
\begin{equation}
\theta=\frac{1}{2}\atanTwo(C,2\Omega),\qquad
\ket{\chi_-}=\cos\theta\,\ket{+,-}+\sin\theta\,\ket{-,+}.
\label{eq:mixing-angle}
\end{equation}
The orthogonal vector $-\sin\theta\ket{+,-}+\cos\theta\ket{-,+}$ has the upper mixed eigenvalue. Equal weights arise asymptotically when $|C|\gg2\Omega$, not generically at nonzero $\Omega$. For $|C|\ll2\Omega$,
\begin{equation}
R=2\Omega+\frac{C^2}{4\Omega}+O(C^4/\Omega^3),\qquad
\delta E_{n,\mathrm{mix},\pm}^2=\pm\frac{C^2}{4\Omega}+O(C^4/\Omega^3).
\label{eq:comparison-expansion}
\end{equation}
Thus weak transverse mixing gives quadratic shifts in $\beta\rho$ at fixed $n$\old{. At} $C=0$, the upper mixed level at $n$ and lower mixed level at $n+1$ coincide; their squared-energy separation at nonzero $C$ is $2R-4\Omega=C^2/(2\Omega)+O(C^4/\Omega^3)$, also quadratic.

At $\rho=0$, simultaneous $\sigma^3,T^3$ eigenstates can be used. For fixed $n,s$ the two isospin radicands differ by $2A$, giving, in the weak aligned regime,
\begin{equation}
\zeta=\frac{g^2\beta^2}{2m},\qquad
|\Delta E_{\mathrm{aligned}}^{(+)}|\simeq\frac{m\zeta}{E_{n,s}^{(0)}}.
\label{eq:aligned-gap}
\end{equation}
Here $E_{n,s}^{(0)}=\sqrt{m^2+2\Omega(2n+1-s)}$ \old{and} $A\ll(E_{n,s}^{(0)})^2$. In contrast, \old{the} pure first-order Pauli model at $\rho=0$ has rigid isospin shifts $\pm r$ and separation $|\kappa|\zeta$. These models share the generator $T^3$, not the same spectrum or spin dependence.

\subsection{Positivity and numerical check}\label{sec:positivity}
\old{For} $m>0,\ \Omega>0$, nonnegativity of the entire comparison operator is equivalent to
\begin{equation}
m^2+2\Omega-\frac{g^2\beta^2}{2}-\sqrt{4\Omega^2+g^4\beta^2\rho^2}\geq0.
\label{eq:positivity}
\end{equation}
It is attained first at $n=0$ in the lower mixed branch. This is a restriction on $\KPR$, not on the real spectrum of the Hermitian first-order problem\old{.}

For $m=g=1$, $\Omega=0.75$, $\beta=1$, $\rho=0.4$, direct diagonalization gives
\begin{equation}
\begin{aligned}
\spec(\left.\KPR\right|_0)&=\{0.4475825304,1.5,3.5524174696,4.5\},\\
E^{(+)}&=\{0.6690160913,1.2247448714,1.8847857888,2.1213203436\}.
\end{aligned}
\label{eq:comparison-numerical}
\end{equation}
At fixed $\rho=0.4$ \old{the} positive-$\beta$ boundary in Eq.~\eqref{eq:positivity} is $\beta_{\max}=1.34629947$. Figure~\ref{fig:comparison} uses the four-branch spectrum and stays inside this domain.

\begin{figure}[tbp]
\centering
\includegraphics[width=0.91\textwidth]{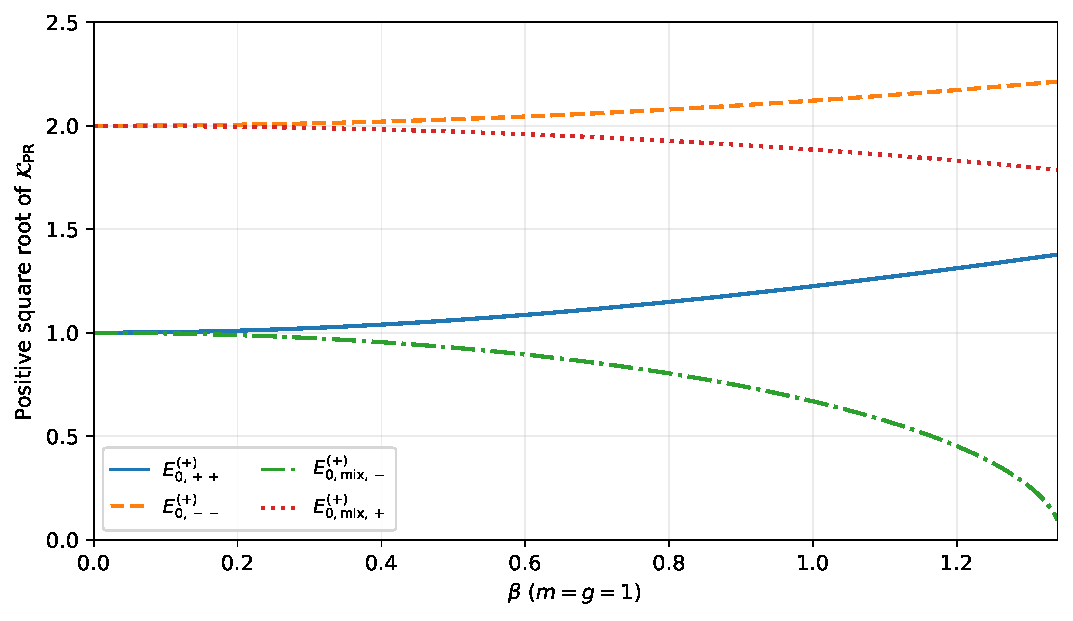}
\caption{\rev{Positive square roots of $\left.\KPR\right|_{n=0}$ for $m=g=1$, $\Omega=0.75$, $\rho=0.4$. There are four internal branches, not an additional independent spin label for every internal eigenvalue. The two aligned states are not degenerate at fixed $n$. The plotted interval ends below the positivity boundary $\beta_{\max}=1.34629947$. This is a comparison-model plot, not the spectrum of $\Hmin$.}}
\label{fig:comparison}
\end{figure}
\FloatBarrier
\end{refereechange}

\begin{refereechange}
\section{Gauge invariants and internal-basis labels}\label{sec:gauge}
\subsection{Background invariants and the meaning of the parameters}\label{sec:invariants}
Write $\widehat F_{\mu\nu}=F_{\mu\nu}^aT^a$. Under a local internal transformation $U(x,t)$ with $\Psi'=U\Psi$, the connection and curvature transform as
\begin{equation}
\mathcal{A}'_\mu=U\mathcal{A}_\mu U^\dagger+\ii(\partial_\mu U)U^\dagger,
\qquad\widehat F'_{\mu\nu}=U\widehat F_{\mu\nu}U^\dagger.
\label{eq:gauge-transformation}
\end{equation}
The quadratic internal invariants for the prescribed background are
\begin{equation}
2\tr(\widehat F_{12}^2)=g^2\beta^4,\qquad
2\tr(\widehat F_{01}^2+\widehat F_{02}^2)=2g^2\beta^2\rho^2.
\label{eq:quadratic-invariants}
\end{equation}
For fixed $g$ they determine $\beta^2$ and $|\beta\rho|$. The constant transformation $U_3=e^{\ii\pi T^3}$ sends $(\beta,\rho)$ to $(-\beta,\rho)$ within the same ansatz. Thus the sign of $\beta$, and hence the sign of $\beta\rho$ for a fixed $\rho$, is not gauge invariant.

These quadratic contractions do not completely classify a connection. In particular, one cannot conclude that $\rho$ has invariant content only through $|\beta\rho|$. Define the adjoint derivative
\begin{equation}
D_0\widehat F=\partial_0\widehat F+\ii[g\rho T^3,\widehat F].
\label{eq:adjoint-derivative}
\end{equation}
For the static ansatz,
\begin{equation}
D_0\widehat F_{02}=-g^2\beta\rho^2T^1,\qquad
\mathcal{J}=2\tr(\widehat F_{01}D_0\widehat F_{02})=g^3\beta^2\rho^3.
\label{eq:cubic-invariant}
\end{equation}
Consequently,
\begin{equation}
\frac{\mathcal{J}}{2\tr(\widehat F_{01}^2)}=g\rho,\qquad\beta\rho\ne0.
\label{eq:signed-rho}
\end{equation}
Both numerator and denominator are invariant under internal gauge transformations. They are specified spacetime components, not Lorentz scalars; the coordinate orientation is held fixed. Thus opposite signed $\rho$ values in the canonical static ansatz are not generally gauge equivalent. This statement concerns the classification of that ansatz, not the gauge invariance of an arbitrary individual $A_0$ component.

At $\beta=0$ the curvature vanishes, and $A_0=\rho T^3$ is removable locally by $U(t)=e^{\ii g\rho tT^3}$, provided no temporal boundary condition obstructs that transformation. At $\beta\ne0$ the same transformation makes the spatial connection time dependent. Even at $\rho=0$, the present spatial connection has $D_i\widehat F_{12}\ne0$, unlike an Abelian-embedded connection with the same constant magnetic curvature and vanishing covariant derivative. This illustrates why a field-strength value alone does not determine a non-Abelian potential \cite{WuYang1975}.

\subsection{Stationary spectra and the covariance of each model}\label{sec:covariance}
For the full minimally coupled Hamiltonian, with or without the covariant Pauli term, time-independent gauge transformations give $H'=UHU^\dagger$ on correspondingly transformed domains. They preserve the stationary spectrum. Under a time-dependent transformation,
\begin{equation}
H'=UHU^\dagger+\ii\dot U U^\dagger.
\label{eq:time-gauge}
\end{equation}
Stationary energy labels therefore require a fixed time-translation/gauge convention. Physical predictions remain covariant when states, observables, and probes are transformed together; one must not compare the raw stationary eigenvalues of two differently time-dependent frames as if they were unchanged.

The internal diagnostic $\LNA$ is pointwise covariant, and its unordered eigenvalue set is invariant. The full $\KPR$, however, contains only Abelian orbital derivatives. Spatial derivatives of a nonconstant $U$ are not compensated by a non-Abelian kinetic connection. Thus $\KPR$ as defined has global internal-basis covariance, not an independently established local $\mathrm{SU}(2)$ gauge covariance. The same qualification applies to the fixed-frame pure Pauli comparison $\HB+V_P$ when the non-Abelian minimal connection is omitted. Retaining a locally gauged fundamental doublet would require the appropriate covariant kinetic terms and would change that model\old{.}

\subsection{What ``aligned'', ``singlet'', and ``triplet'' mean}\label{sec:labels}
The isolated internal spectrum can be written invariantly as the aligned eigenvalue $g^2\beta^2/(4m)$ with multiplicity two, and the unordered pair
\begin{equation}
\left\{-\frac{g^2\beta^2}{4m}+\frac{g^2|\beta\rho|}{2m},\quad
-\frac{g^2\beta^2}{4m}-\frac{g^2|\beta\rho|}{2m}\right\}.
\label{eq:unordered-pair}
\end{equation}
The transformation $U_3$ sends $\ket{S}\mapsto-\ii\ket{T_0}$ \old{and} $\ket{T_0}\mapsto-\ii\ket{S}$, while mapping the background to $-\beta$. The names assigned to these basis combinations are not separately gauge-invariant physical labels. Their isolated eigenvalue separation has invariant magnitude $g^2|\beta\rho|/m$, but it is not a universal energy gap.

For nonzero magnetic curvature, $\hat n^a=F_{12}^a/\sqrt{F_{12}^bF_{12}^b}$ defines a covariant internal axis. In the present convention $\hat n=-\hat e_3$, and the projector onto the aligned diagnostic subspace is $P_\parallel=\tfrac12\mathbb{I}-\sigma^3\hat n^aT^a$. This geometric definition transforms with \old{the} background. It is undefined at $\beta=0$ and does not make ``aligned'' and ``mixed'' conserved quantum numbers of the entire first-order Hamiltonian: $\HB$ acts on the spinor factor. Likewise, the equal weights in Eq.~\eqref{eq:singlet-triplet} describe $\LNA$ in an adapted basis, whereas the full $\KPR$ has the generally unequal weights of Eq.~\eqref{eq:mixing-angle}. The first-order eigenstates instead involve the distinct oscillator components in Eq.~\eqref{eq:basis}.\label{sec:gauge-end}
\end{refereechange}

\begin{refereechange}
\section{Conditional regimes for resolving the first-order splitting}\label{sec:regimes}
\subsection{Energy conventions and a resolution criterion}\label{sec:resolution}
For an effective Dirac velocity $v$, denote the mass energy by $\mathcal{M}$ and the \old{two} minimal-coupling energy coefficients by
\begin{equation}
U_\beta=vg\beta,\qquad U_\rho=g\rho,\qquad r=U_\rho/2,\qquad k=U_\beta.
\label{eq:energy-units}
\end{equation}
Here the spatial connection is expressed in momentum units and the temporal connection in energy units. These equations concern minimal coupling only; a synthetic minimal gauge potential does not automatically implement an anomalous Pauli interaction. In a magnetic realization with no oscillator contribution,
\begin{equation}
\varepsilon_q=\sqrt{\mathcal{M}^2+2q\hbar eBv^2},\qquad
\ell_B=\sqrt{\hbar/(eB)},\qquad eB>0.
\label{eq:magnetic-units}
\end{equation}
The oscillator formulas apply after the corresponding energy-squared scale is matched. Massless examples below concern the minimal model, never a $1/m$ Pauli coupling\old{.}

A spectroscopic avoided crossing should be compared with a specified probe and a broadening budget. A sufficient target is
\begin{equation}
\Delta_q\gg\Gamma_L,\quad\Delta_q\gg k_BT,\qquad |k|\ll\Drem
\label{eq:resolution}
\end{equation}
when the leading two-level expression is used. Here $\Gamma_L$ is an energy linewidth, not a frequency; the temperature condition is a conservative criterion for thermally broadened spectroscopy. Suitable state preparation and nonzero probe matrix elements are also required. Detuning should be swept across the crossing; the complete blocks remain valid outside the two-level approximation.

\subsection{A graphene-inspired example and its limitations}\label{sec:graphene}
Graphene continuum descriptions naturally admit effective gauge fields \cite{CastroNeto2009,Vozmediano2010}, and spatially modulated interlayer hopping in sheared or twisted bilayers can enter as non-Abelian gauge potentials \cite{SanJose2012}. This motivates, but does not establish, a constant-connection approximation. The required four-component identification is explicit: $\sigma$ acts on the two sublattices and $\tau$ on the two layers, with real spin and the second valley treated as spectators when their mixing is negligible. After a chosen basis/orientation transformation, the required local Hamiltonian is
\begin{equation}
H_{\mathrm{target}}=v(\sigma^1\Pi_x-\sigma^2\Pi_y)+\mathcal{M}\sigma^3
+\frac{U_\rho}{2}\tau^3-\frac{U_\beta}{2}(\sigma^1\tau^2+\sigma^2\tau^1).
\label{eq:target}
\end{equation}
Thus $U_\rho$ is a layer on-site-energy difference, not automatically a measured band gap, and the coefficient of each displayed transverse Pauli product is $U_\beta/2$. A microscopic device mapping must establish these coefficients and quantify all extra tunnelling, scalar, warping, disorder, and intervalley terms.

Consider the conditional parameter choice $\mathcal{M}=0$, $q=1$, $\varepsilon_1=36\ \mathrm{meV}$, and $U_\beta=10\ \mathrm{meV}$. The bare resonance is at $U_\rho=36(\sqrt{2}-1)=14.911688\ \mathrm{meV}$. The exact four signed roots of the block, in meV, are
\begin{equation}
\{-59.43080532,-28.07162061,38.75197959,48.75044634\}.
\label{eq:graphene-roots}
\end{equation}
The positive-pair gap is $9.998467\ \mathrm{meV}$, close to the leading $10\ \mathrm{meV}$. For example, $T\leq4\ \mathrm{K}$ gives $k_BT\leq0.345\ \mathrm{meV}$; requiring $\Gamma_L\leq1\ \mathrm{meV}$ would leave an order-of-magnitude linewidth margin. These are design requirements, not linewidths measured in a device described here.

Taking $v=10^6$~m/s as an illustrative velocity, the chosen $\varepsilon_1$ corresponds to a field of approximately $1\ \mathrm{T}$. A $q=1$ cyclotron wave packet samples a radius of order $\sqrt{5}\ell_B\simeq60\ \mathrm{nm}$. The \old{non-Abelian} coefficients must be approximately constant over that region, requiring a variation length $L_{\mathrm{var}}\gg60\ \mathrm{nm}$ as well as the amplitude matching in Eq.~\eqref{eq:target}. A generic short-period moir\'e pattern does not automatically satisfy this requirement. The example is therefore a reproducible spectral benchmark and an explicit experimental target, not a demonstrated prediction for an arbitrary bilayer device.

\subsection{Cold-atom and ion-simulator targets}\label{sec:atoms}
Laser-dressed atoms provide experimentally established artificial-gauge-field ingredients \cite{Dalibard2011,Goldman2014}, but their internal spaces must not be conflated. The experiment of Lin et al.~\cite{Lin2011} realizes equal Rashba and Dresselhaus strengths, rather than by itself establishing the two independent noncommuting spatial components required here. Two-dimensional spin--orbit coupling in optical Raman lattices \cite{Wu2016} and tunable Dirac points in honeycomb lattices \cite{Tarruell2012} provide complementary ingredients. They do not alone demonstrate the complete four-component oscillator Hamiltonian.

A possible target uses a sublattice or band doublet for $\sigma$ and a separate atomic doublet for $\tau$. The effective mass and velocity, two transverse spinor--isospin matrix elements, detuning $U_\rho$, and the orbital magnetic/oscillator ladder must all be engineered. One cannot use the same two atomic states as two independent tensor factors. Interaction, trap, and laser-induced broadening must be included in the resolution budget.

For a numerical target in recoil units, take the massless minimal block with $\varepsilon_1=2E_R$, $U_\beta=0.25E_R$\old{, and} $U_\rho=2(\sqrt{2}-1)E_R=0.828427E_R$. Its exact positive roots are $2.29254937E_R$ \old{and} $2.54254154E_R$, so
\begin{equation}
\Delta=0.249992E_R,\qquad\frac{E_R}{h}=\frac{h}{2M_{\mathrm{at}}\lambda_L^2}\simeq3678.38\ \mathrm{Hz}
\quad({}^{87}\mathrm{Rb},\ \lambda_L=790\ \mathrm{nm}).
\label{eq:recoil-example}
\end{equation}
Here $\lambda_L$ defines the chosen recoil unit. The predicted model gap is $\Delta/h=919.57\ \mathrm{Hz}$ and $\Delta/k_B\simeq44.1\ \mathrm{nK}$. A linewidth target $\Gamma_L/h\leq100\ \mathrm{Hz}$ and an effective thermal scale $T\lesssim5\ \mathrm{nK}$ would provide useful separation; a transform-limited interrogation would also require a duration of order $10\ \mathrm{ms}$ or longer for a $100\ \mathrm{Hz}$ frequency resolution. These are demanding conditional targets, not reported performance of the cited experiments. State-prepared nonequilibrium spectroscopy may have a different thermal requirement, which must be evaluated for its actual protocol.

The Dirac-oscillator/Jaynes--Cummings correspondence \cite{Bermudez2007} also motivates an ion simulator. In ladder form, $-k\Gamma=\ii k(\sigma^+T^+-\sigma^-T^-)$ is a simultaneous spinor--isospin flip, in addition to the oscillator sideband and longitudinal field. Its implementation requires a second independent doublet or a suitable four-level encoding; the one-doublet Dirac simulation of Ref.~\cite{Gerritsma2010} is not by itself the full system. We make no quantitative trapped-ion gap or feasibility claim without a specified coupling and coherence budget\old{.}

The experimentally meaningful objective is a tunable, resolvable avoided crossing derived from Eq.~\eqref{eq:quartic}. The product-linear resonant signature belongs to the pure Pauli model with a separately engineered $V_P$; the two numerical platform targets above concern minimal coupling and have $k=U_\beta$.\label{sec:regimes-end}
\end{refereechange}

\section{Conclusion}\label{sec:conclusion}
The uniform \old{non-Abelian} connection produces a chromomagnetic curvature along $T^3$ and transverse chromoelectric curvatures proportional to $\beta\rho$. The exact first-order Hamiltonian nevertheless depends on the connection, not just its curvature. Its conserved $Q$ sectors give an exact quartic for $q\geq1$, a cubic for $q=0$, and one isolated signed level. This structure makes the exceptional lowest states explicit and avoids extraneous signs introduced by squaring.

Minimal coupling and pure Pauli coupling have different physical content. The former already mixes spinor and isospin through $g\beta$ when $\rho=0$; the latter has an aligned internal-Zeeman term generated by $T^3$ and a transverse coefficient proportional to $\beta\rho$. A gap linear in that coefficient appears near a tuned resonance, while nonresonant shifts are quadratic. Equal-weight mixing is a leading two-level statement, not an exact finite-coupling assertion.

The isolated field-strength diagnostic retains its simple aligned and singlet/triplet eigenvalues. The complete phenomenological operator $\KPR$ does not inherit their additive spectrum because its backbone spin term does not commute with the transverse coupling. Its four branches, unequal-weight mixed states, and positivity condition are stated separately from the Dirac spectrum. Gauge-invariant curvatures and covariant-derivative contractions are likewise distinguished from basis-dependent state names and stationary-frame energy labels\old{.}

Conditional meV- and recoil-scale examples show what spectral resolution a proposed implementation would require. A microscopic mapping, spatial-uniformity regime, probe matrix elements, and measured broadening remain necessary before either example can be interpreted as a device-specific prediction. Extensions to larger isospin, spatially varying backgrounds, and explicit simulator protocols can build on the finite-block results established here.

\appendix
\begin{refereechange}
\section{Formal second-order reduction of the minimal Dirac equation}\label{app:reduction}
\subsection{Kinetic operators and all commutator terms}\label{app:kinetic}
Define
\begin{equation}
K_1=\Pi_1-g\beta T^2,\quad K_2=\Pi_2+g\beta T^1,\quad
D=\sigma^1K_1-\sigma^2K_2+m\sigma^3,\quad\mathcal{A}_0=g\rho T^3.
\label{eq:kinetic-operators}
\end{equation}
Then $\Hmin=D+\mathcal{A}_0$, and
\begin{align}
\old{[}K_1,K_2\old{]}&=-\ii(2\Omega-g^2\beta^2T^3),\label{eq:kinetic-commutator}\\
\old{[}\mathcal{A}_0,K_j\old{]}&=\ii g^2\beta\rho T^j\quad(j=1,2),\notag\\
\old{[}\mathcal{A}_0,D\old{]}&=\ii g^2\beta\rho(\sigma^1T^1-\sigma^2T^2).
\label{eq:electric-commutator}
\end{align}
\old{The} first-order eigenvalue equation reads $(E-\mathcal{A}_0)\psi=D\psi$. Applying $E-\mathcal{A}_0$ and respecting operator order gives
\begin{equation}
(E-\mathcal{A}_0)^2\psi=\bigl(D^2-\old{[}\mathcal{A}_0,D\old{]}\bigr)\psi.
\label{eq:stationary-identity}
\end{equation}
No commutation between the temporal potential and the kinetic Dirac operator has been assumed. The square of $D$ is
\begin{equation}
\begin{aligned}
D^2&=K_1^2+K_2^2+m^2-\ii\sigma^3[K_1,K_2]\\
&=\Pi_1^2+\Pi_2^2+m^2-2\Omega\sigma^3+g^2\beta^2\sigma^3T^3
+\frac{g^2\beta^2}{2}-2g\beta(\Pi_1T^2-\Pi_2T^1).
\end{aligned}
\label{eq:d-square}
\end{equation}
The exact stationary second-order identity is therefore
\begin{equation}
\begin{aligned}
\left[E^2-2g\rho ET^3+\frac{g^2\rho^2}{4}\right]\psi
={}&\Bigl[\Pi_1^2+\Pi_2^2+m^2-2\Omega\sigma^3+g^2\beta^2\sigma^3T^3+\frac{g^2\beta^2}{2}\\
&\qquad-2g\beta(\Pi_1T^2-\Pi_2T^1)-\ii g^2\beta\rho(\sigma^1T^1-\sigma^2T^2)\Bigr]\psi.
\end{aligned}
\label{eq:full-second-order}
\end{equation}
The magnetic and electric field-strength terms are respectively $-g\sigma^3F_{12}^aT^a$ and $\ii g\alpha^iF_{0i}^aT^a$. The anti-Hermitian electric term in Eq.~\eqref{eq:full-second-order} is not a defect in the Hermiticity of $\Hmin$: this equation is an energy-dependent stationary identity, not an assertion that its right-hand side is the ordinary Hamiltonian square.

For completeness, the actual square is manifestly Hermitian:
\begin{equation}
\begin{aligned}
H_{\mathrm{min}}^2={}&\Pi_1^2+\Pi_2^2+m^2-2\Omega\sigma^3+g^2\beta^2\sigma^3T^3
+\frac{g^2\beta^2}{2}+\frac{g^2\rho^2}{4}\\
&-2g\beta(\Pi_1T^2-\Pi_2T^1)+2g\rho\HB T^3.
\end{aligned}
\label{eq:hmin-square}
\end{equation}
Here $[\HB,T^3]=0$ and $\{\Gamma,T^3\}=0$. Every first-order eigenstate satisfies Eq.~\eqref{eq:full-second-order}, but a solution of a squared relation does not automatically supply an admissible signed first-order state. The reconstruction is fixed without ambiguity by the blocks of Sec.~\ref{sec:blocks}.

\subsection{Why the comparison model is not an exact reduction}\label{app:comparison}
Relative to \old{Eq.}~\eqref{eq:full-second-order}, $\KPR$ retains the Abelian backbone and the chromomagnetic term $g^2\beta^2\sigma^3T^3$. It omits the isospin--orbit term $-2g\beta(\Pi_1T^2-\Pi_2T^1)$, the diamagnetic constant $g^2\beta^2/2$, and the scalar-potential contributions $2g\rho ET^3-g^2\rho^2/4$. It also replaces the electric commutator by the different Hermitian operator $-g^2\beta\rho(\sigma^1T^1+\sigma^2T^2)$.

The isospin--orbit term is linear in $\beta$, connects adjacent oscillator states, and gives second-order effects of the same $\beta^2$ order as the retained magnetic term when nondegenerate perturbation theory applies. The diamagnetic term is already of that order. There is no stated hierarchy under which these omissions are all parametrically smaller. Furthermore, the exact electric commutator connects the $\ket{+,+},\ket{-,-}$ pair, whereas the Hermitian comparison term connects the $\ket{+,-},\ket{-,+}$ pair. The replacement changes the operator, not just its notation. These facts justify treating $\KPR$ only as the separate model solved in Sec.~\ref{sec:full-comparison}.\label{app:reduction-end}
\end{refereechange}

\section{Derivation of the secular equation and resonant projection}\label{app:secular}
For a vector $(x_1,x_2,x_3,x_4)^T$ in the phase-adjusted $q\geq1$ basis, the third and second rows of $H_qx=Ex$ give
\begin{equation}
(E-r+m)x_3=a_qx_1,\qquad (E+r-m)x_2=a_{q+1}x_4.
\label{eq:eliminate}
\end{equation}
Substitution in the remaining rows yields
\begin{equation}
\begin{aligned}
\bigl[(E-r)^2-\varepsilon_q^2\bigr]x_1&=k(E-r+m)x_4,\\
\bigl[(E+r)^2-\varepsilon_{q+1}^2\bigr]x_4&=k(E+r-m)x_1.
\end{aligned}
\label{eq:remaining-rows}
\end{equation}
Their determinant gives Eq.~\eqref{eq:quartic}. Although this elimination is convenient away from vanishing denominators, Eq.~\eqref{eq:quartic} is a polynomial identity for $\det(E\mathbb{I}-H_q)$ and remains valid at those special values. No roots should be lost by dividing by an energy-dependent factor. Removing the absent third basis vector gives Eq.~\eqref{eq:cubic}. At $r=0$ this cubic factors as
\begin{equation}
(E-m)(E^2-m^2-4\Omega-k^2)=0,
\label{eq:cubic-factor}
\end{equation}
which establishes the signed lowest-sector qualification directly.

For the nonresonant shift, write the quartic as $f(E)-k^2g(E)=0$ and expand around the simple root $E_0=r+\varepsilon_q$. Since
\begin{equation}
\begin{aligned}
f'(E_0)&=2\varepsilon_q[(\varepsilon_q+2r)^2-\varepsilon_{q+1}^2],\\
g(E_0)&=(\varepsilon_q+m)(\varepsilon_q+2r-m),
\end{aligned}
\label{eq:fg}
\end{equation}
$\delta E=k^2g(E_0)/f'(E_0)+O(k^4)$ gives Eq.~\eqref{eq:shift}. Its denominator explicitly shows why this expansion fails at a resonance.

The positive eigenvector of $\left(\begin{smallmatrix}m&a_q\\a_q&-m\end{smallmatrix}\right)$ has upper and lower weights $\sqrt{(\varepsilon_q+m)/(2\varepsilon_q)}$ and $\sqrt{(\varepsilon_q-m)/(2\varepsilon_q)}$. The perturbation $k$ in $H_q$ connects the first component of the isospin-up doublet to the fourth component of the isospin-down doublet. Its matrix element is exactly Eq.~\eqref{eq:mq}. To leading order the positive-level projection is
\begin{equation}
H_{\mathrm{eff}}=\begin{pmatrix}
r+\varepsilon_q&M_q\\M_q&-r+\varepsilon_{q+1}
\end{pmatrix}+O(k^2/\Drem),
\label{eq:effective-h}
\end{equation}
whose separation is $\sqrt{(2r+\varepsilon_q-\varepsilon_{q+1})^2+4|M_q|^2}$ before remote-state corrections. This proves the gap formula and \old{the} limited status of equal-weight mixing at the bare resonance.

\begin{refereechange}
\section*{Data availability}
No new experimental data were generated. The numerical values used in the figures and benchmark examples are determined by the equations and parameters stated in the manuscript\old{.}
\end{refereechange}
\end{revised}


\end{document}